\documentclass[prl,showpacs,floatfix,amsmath,amssymb,superscriptaddress]{revtex4}
\usepackage{latexsym}
\usepackage{graphicx}
\usepackage{times}
\usepackage{amsmath}
\usepackage{dcolumn}
\usepackage{subfigure}
\usepackage{latexsym,amsmath,amssymb,bm,euscript}
\def\MnScS{$\text{Mn}\text{Sc}_2\text{S}_4$\,}

\def\CoAl{$\text{Co}\text{Al}_2\text{O}_4$\,}

\def\DM{\text{Dzyaloshinskii-Moriya\ }} 
\def\Neel{\text{N\'eel\ }}

\newcommand{\be}{\begin{equation}}
\newcommand{\ee}{\end{equation}}

\begin{document}

\title{Order by disorder and spiral spin liquid in\\
frustrated diamond lattice antiferromagnets} 
\author{Doron Bergman$^*$}
\affiliation{Department of Physics, University of California, Santa Barbara, CA
93106-9530}

\author{Jason Alicea}
\affiliation{Department of Physics, University of California, Santa Barbara, CA
93106-9530}

\author{Emanuel Gull}
\affiliation{Theoretische Physik, Eidgen\"ossische Technische Hochschule Z\"urich, 
CH-8093 Z\"urich, Switzerland}

\author{Simon Trebst}
\affiliation{Microsoft Research, Station Q, University of California, Santa Barbara, CA 93106}

\author{Leon Balents}
\affiliation{Department of Physics, University of California, Santa Barbara, CA 93106-9530}

\date{\today} 

\begin{abstract} {\sl Frustration} refers to competition between
different interactions that cannot be simultaneously satisfied, a
familiar feature in many magnetic solids.  Strong frustration results
in highly degenerate ground states, and a large suppression of
ordering by fluctuations.  Key challenges in frustrated magnetism are
characterizing the fluctuating {\sl spin-liquid} regime and
determining the mechanism of eventual order at lower temperature.
Here, we study a model of a diamond lattice antiferromagnet
appropriate for numerous spinel materials.  With
sufficiently strong frustration a massive ground state degeneracy
develops amongst spirals whose propagation wavevectors reside
on a continuous two-dimensional ``spiral surface'' in momentum space.  We argue
that an important ordering mechanism is {\sl entropic} splitting of
the degenerate ground states, an elusive phenomena called 
{\sl order-by-disorder}.  A broad ``spiral spin-liquid'' regime emerges at
higher temperatures, where the underlying spiral surface can be directly
revealed via spin correlations.  We discuss the agreement
between these predictions and the well characterized spinel \MnScS.
\end{abstract}

\maketitle 


When microscopic interactions in a material conspire to ``accidentally''
produce many nearly degenerate low-energy states, otherwise weak
residual effects can give rise to remarkable emergent behavior.  This
theme recurs throughout modern condensed matter physics.  Quintessential
examples include the cuprates, with several competing orders including
high-T$_c$ superconductivity, and exotic quantum (Hall) liquids in
two-dimensional electron systems, arising from partial Landau-level
occupation.  Insulating magnets
constitute a particularly abundant source of such phenomena, as in
numerous cases {\sl frustration} generated by the competition between
different exchange interactions leads to large classical ground-state
degeneracies.  An important experimental signature of such degeneracies
is an anomalously low ordering temperature $T_c$ relative to the Curie
Weiss temperature $\Theta_{CW}$; indeed, values of the 
``frustration parameter'' $f = |\Theta_{CW}|/T_c$ larger than 5-10 are
typically taken as empirical evidence of a highly frustrated magnet.\cite{ramirez94:_stron_geomet_frust_magnet}
This sharp suppression of $T_c$ opens up a broad 
``spin-liquid'' regime for temperatures $T_c \lesssim T \lesssim
|\Theta_{CW}|$, where the system fluctuates amongst the many low-energy
configurations but evades long-range order.  Highly non-trivial physics
can emerge here, as attested for instance in
pyrochlore antiferromagnets by the experimental observation of hexagonal
loop correlations in neutron scattering on the spinel 
ZnCr$_2$O$_4$\cite{Lee:nat02}, and theoretically by the establishment of
``dipolar'' correlations.\cite{isakov:167204}\

Low-temperature ordering in highly frustrated magnets
often displays an exquisite sensitivity to degeneracy-breaking
perturbations, notably dipolar interactions and minimal disorder in the
spin-ice pyrochlores,\cite{StevenT.Bramwell11162001}, spin-lattice
coupling in various spinels\cite{yamashita:4960}, and \DM interactions
in Cs$_2$CuCl$_4$\cite{veillette:214426}. 
However, the lifting of degeneracy need not
require the presence of such explicit perturbations.  This can be achieved,
rather remarkably, by \emph{fluctuations} -- a process commonly referred
to as ``order-by-disorder''.\cite{Villain:80} Here, degeneracy in the
free energy is lifted {\sl entropically},
resulting in ordering which counter-intuitively is enhanced by
increasing temperature.  An analogous phenomenon occurs in quantum spin
models at $T = 0$, where quantum fluctuations
provide the degeneracy-breaking mechanism \cite{OldQuantumOBDO,
  Henley:89, Chubukov}.  
Whether or not order-by-disorder transpires depends crucially 
upon the degree of degeneracy: it is known to occur in various FCC 
antiferromagnets\cite{Henley:87,Zhit:05}, for instance, but not in the 
more severely degenerate nearest-neighbor pyrochlore
antiferromagnet\cite{PhysRevB.45.7287,Moessner:prb98}, where instead a
classical spin-liquid regime extends down to $T = 0$.  
While these ideas have
existed for decades and enjoy broad acceptance in the theoretical
community, compelling experimental evidence for order-by-disorder in
even one example is presently lacking.  

\begin{figure} 
  \begin{center} 
    {\resizebox{10cm}{!}{\includegraphics[width=4.0in]{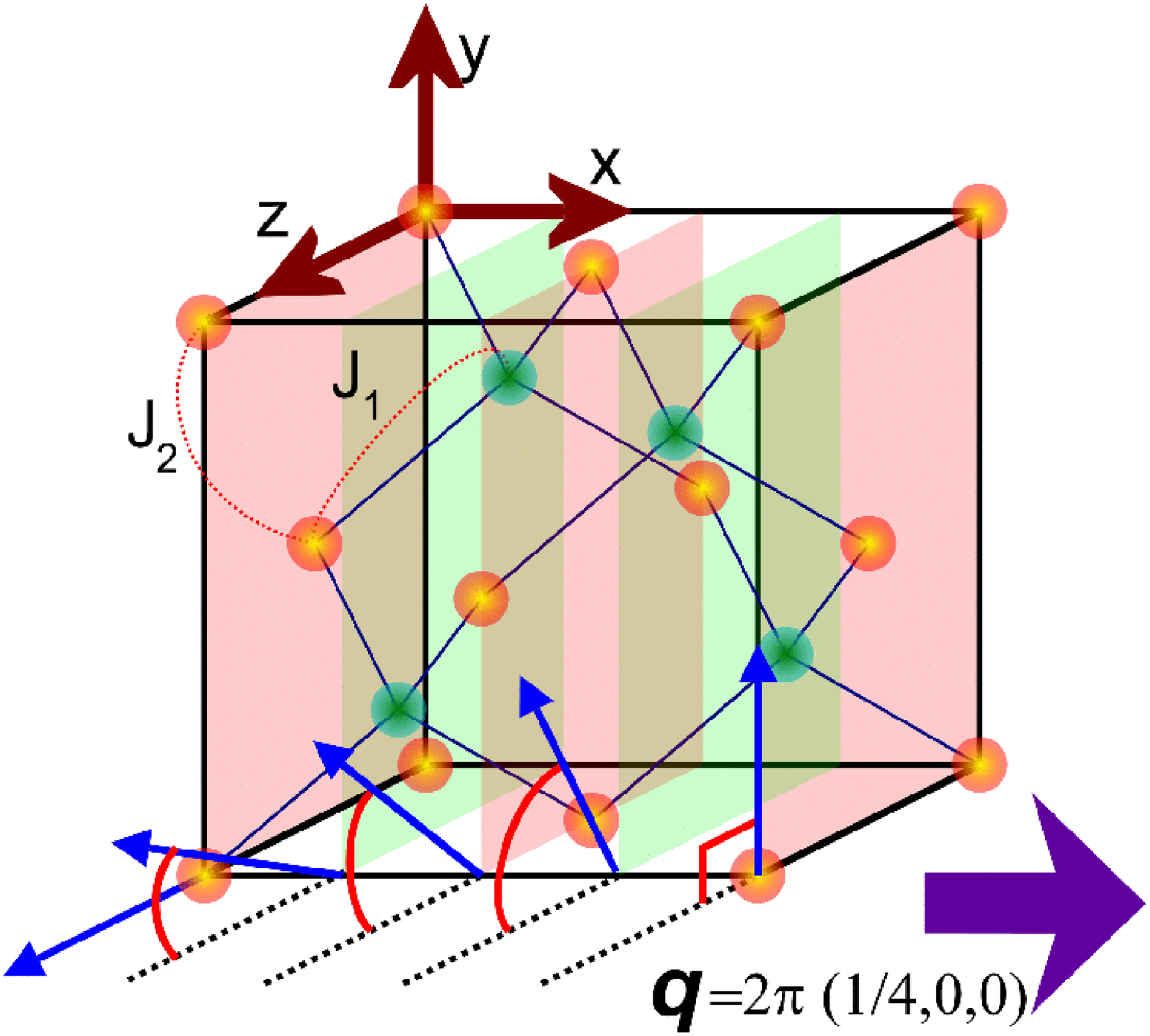}}} 
  \end{center} 
  \caption{The diamond lattice, composed of two interpenetrating FCC
  sublattices (colored orange and green).  Second-neighbor
  antiferromagnetic exchange $J_2$ generates strong frustration that
  is compounded by the competition from the nearest-neighbor
  exchange $J_1$.  
  For $J_2/J_1> 1/8$, this results in a large
  ground state degeneracy consisting of spin spirals whose
  propagation wavevectors lie on a two-dimensional surface in
  momentum space.  The arrows above denote the orientations of spins
  in the shaded planes for one such spiral with wavevector ${\bf q} =
  2\pi(1/4,0,0)$, illustrated for ferromagnetic $J_1$ for clarity.  
} 
  \label{diamond} 
\end{figure}

Here we argue that entropic effects may play a key role in the physics
of insulating normal spinels, with the generic chemical formula
AB$_2$X$_4$, that comprise antiferromagnets on a \emph{diamond} lattice
formed by magnetic, orbitally non-degenerate A sites (see Fig.\
\ref{diamond}).  Numerous strongly frustrated materials in this class
have been recent subjects of intensive experimental study; in
particular, \CoAl and \MnScS for which $f > 10-20$
\cite{Loidl:prb05,Suzuki:06} and $f \approx 10$ \cite{Loidl:prl04},
respectively, are expected to provide ideal test grounds for the physics
we describe.  We introduce a simple classical model for these materials,
consisting of a basic ``parent'' Hamiltonian supplemented by small
corrections, that exhibits complex behavior in accord with numerous experimental
observations.  Remarkably, ground states of the parent theory are
(for most of its phase space) highly degenerate coplanar spirals,
whose propagation wavevectors form a \emph{continuous
  surface} in momentum space.  Within our parent theory,
order-by-disorder occurs with a dramatically suppressed $T_c$ relative
to $\Theta_{CW}$, and above $T_c$ a ``spiral spin liquid'' regime
emerges where the system fluctuations among these degenerate spirals.
While the small corrections (which we describe) inevitably determine
specific ground states at the lowest temperatures, entropy washes these
out at higher temperatures, allowing the spiral spin liquid and/or
order-by-disorder physics inherent to the parent Hamiltonian to become
visible.  This energy-entropy competition is thus manifest as an
interesting multi-stage ordering behavior.
  
Superficially, the strong frustration inherent in materials such as \MnScS
seems rather puzzling.  Indeed, the diamond lattice is bipartite, and
accordingly a model with only nearest-neighbor spin coupling $J_1$,
whether ferromagnetic or antiferromagnetic, exhibits
\emph{no} frustration.  Additional
interactions must therefore be incorporated to account for the observed
frustration.  We first consider the simplest modification that
achieves this, and assume a Hamiltonian with additional
second-neighbor antiferromagnetic exchange $J_2$:
\begin{equation}
  H = J_1\sum_{\langle i j\rangle} {\bf S}_i \cdot {\bf S}_j + 
  J_2\sum_{\langle\langle i j\rangle\rangle} {\bf S}_i \cdot {\bf
  S}_j.
  \label{H}
\end{equation}
Here the spins ${\bf S}_i$ are modeled as classical three-component unit
vectors (absorbing a factor of $S(S+1)$ into the definition of $J_i$),
appropriate to the large spin values ($S=3/2,5/2$) for these materials.
Throughout we set the lattice constant $a = 1$ and consider $J_2>0$
appropriate for antiferromagnetic exchange.  While the sign of $J_1$ can
always be changed by sending ${\bf S}_i\rightarrow -{\bf S}_i$ on one of
the two diamond sublattices, for ease of discussion we will assume
antiferromagnetic $J_1 > 0$ unless specified otherwise.  Additional
interactions such as further-neighbor exchange may also be present, but
will be assumed small and returned to only at the end of the paper.  As
we will see, the parent Hamiltonian Eq.\ (\ref{H}) leads to a rich
theoretical picture which we argue captures the essential physics
operating in these strongly frustrated materials.

To appreciate the frustration in $H$, it is convenient to view the
diamond lattice as composed of two interpenetrating FCC sublattices
(colored orange and green in Fig.\ \ref{diamond}).  From this perspective,
$J_1$ couples the two FCC sublattices, while $J_2$ couples
nearest-neighbors \emph{within} each FCC sublattice.  The FCC 
antiferromagnet is known to be highly
frustrated,\cite{Smart:66} and hence $J_2$ generates strong frustration
which the competition from $J_1$ further enhances.  In fact, as
argued long ago \cite{Roth:64} and emphasized in
\cite{Loidl:prl04,Loidl:prb06}, due to the similarity in exchange paths
coupling first- and second-neighbor sites in such materials, $J_1$ and
$J_2$ are generally expected to have comparable strengths.  

\begin{figure}[hbt]
	\centering
	\subfigure[]{
	  \label{fig:surface_1}
    \includegraphics[height=2.0in]{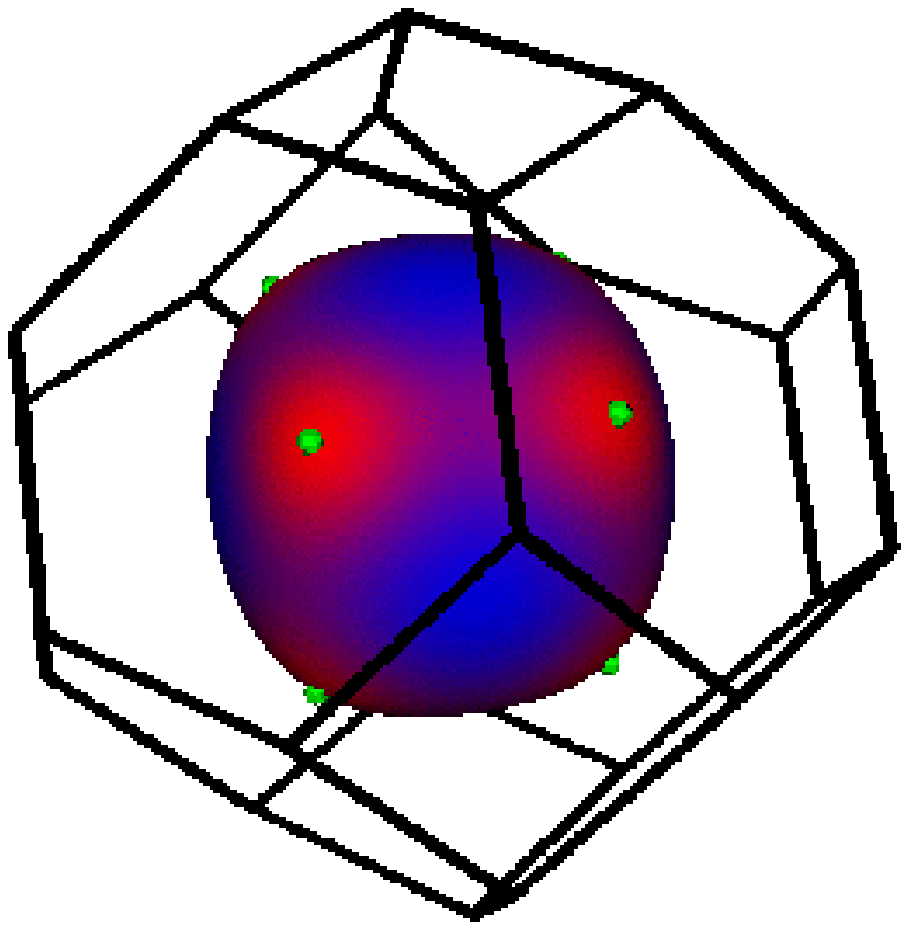}}
	\subfigure[]{
	  \label{fig:surface_2}
    \includegraphics[height=2.0in]{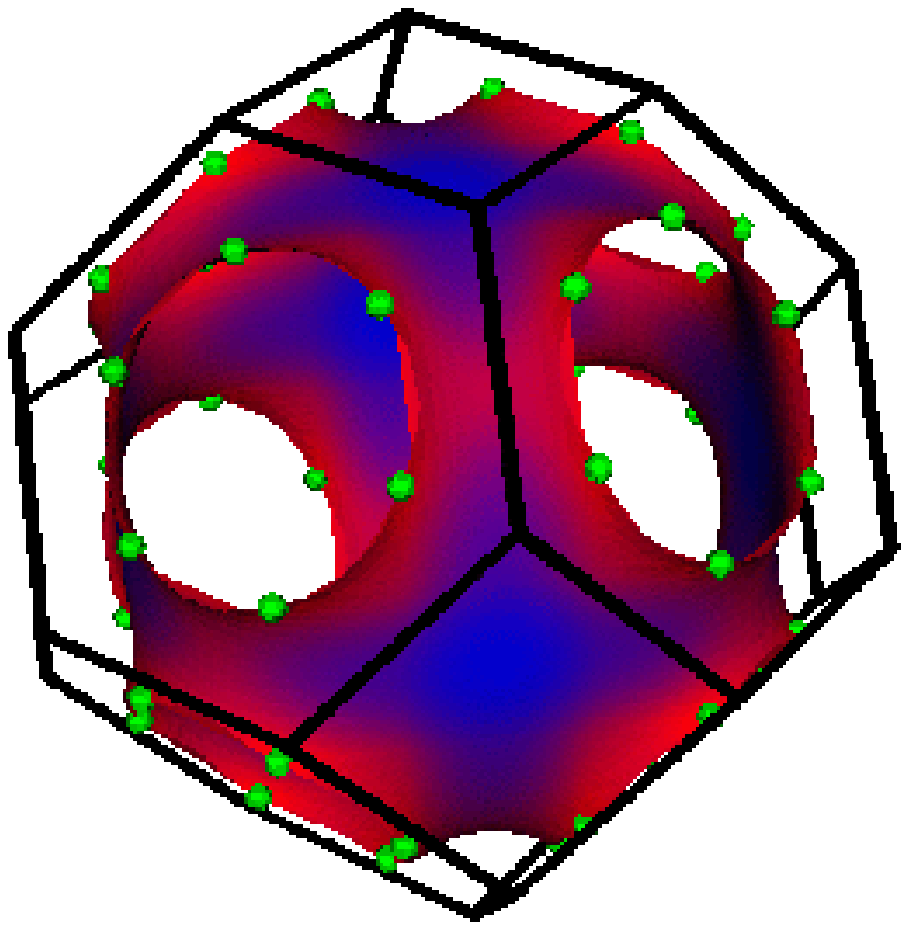}}
    	\subfigure[]{
	  \label{fig:surface_4}
    \includegraphics[height=2.0in]{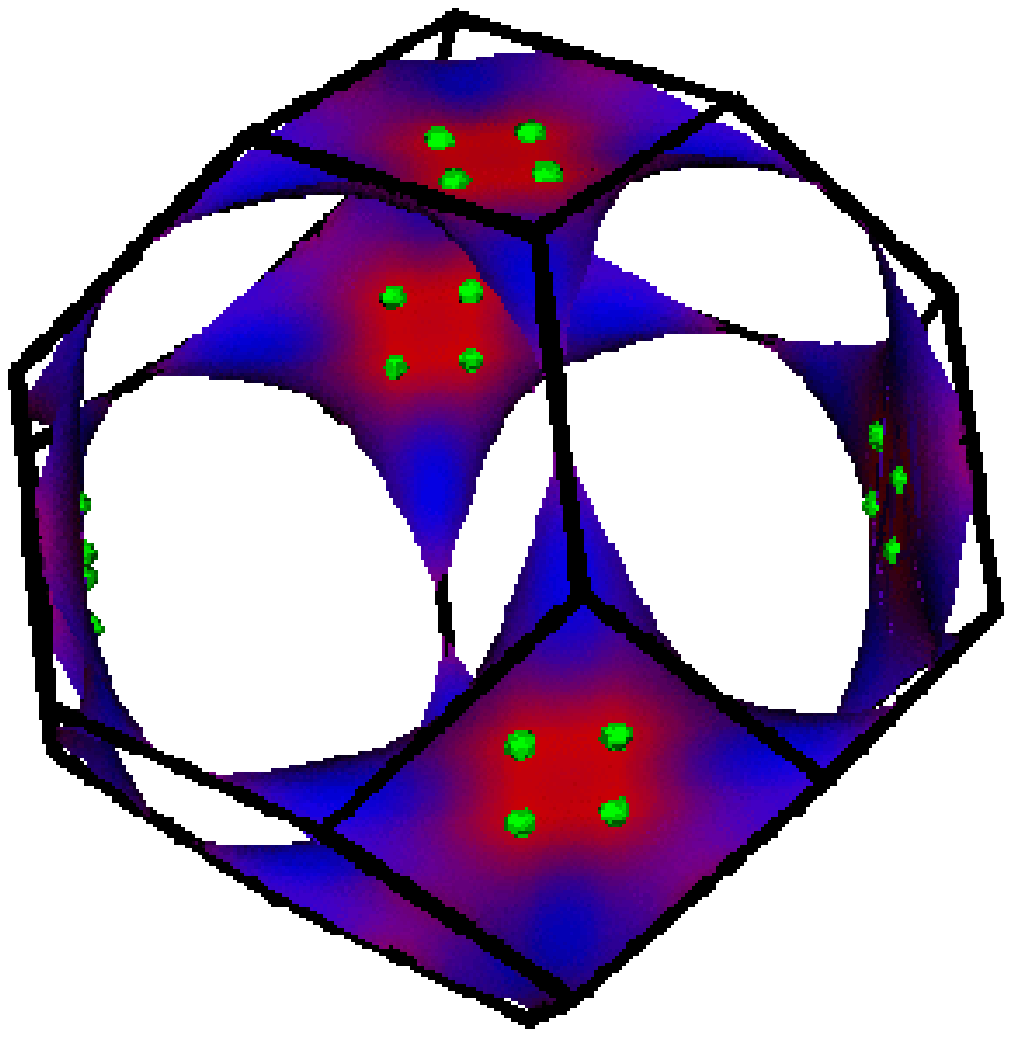}}
   	\caption{``Spiral surfaces'' comprising the degenerate 
    spiral ground state wavevectors for
    coupling strengths $J_2/J_1$ of (a) 0.2, (b) 0.4, and (c)
    0.85, where the last value is appropriate for \MnScS.
    Order-by-disorder occurs at finite temperature, as 
    thermal fluctuations lift the degeneracy in the free energy.  The
    surfaces are color-coded according to the resulting low-temperature 
    free energy at each wavevector, with high values blue, low values
    red, and green the absolute minima.}
	\label{surfaces}
\end{figure}

We begin by discussing the zero-temperature properties of Eq.\ \eqref{H}.
Exact ground states can be obtained for arbitrary
$J_2/J_1$.  In the weakly frustrated
limit with $0 \leq J_2/J_1 \leq 1/8$, the ground state is the \Neel phase, 
with each spin anti-aligned with those of 
its nearest neighbors. For larger
$J_2$ the simple \Neel phase is supplanted by a massively degenerate
set of coplanar spin spirals.  As described above, and illustrated
schematically in Fig.~\ref{diamond}, each spiral ground state
is characterized by a single wavevector ${\bf q}$ lying on a
two-dimensional ``spiral surface''.  
This surface possesses a nearly spherical geometry for 
coupling strengths $1/8 < J_2/J_1 < 1/4$, and exhibits an open
topology for $J_2/J_1 > 1/4$ where it develops ``holes'' centered around the 
(111)-directions (see Fig.\ \ref{surfaces}).  
In the limit $J_2/J_1 \rightarrow \infty$, the surface collapses into 
one-dimensional lines, which are known to characterize the ground
states of the nearest-neighbor-coupled FCC antiferromagnet.\cite{Smart:66}


At small but 
non-zero temperature, one must consider both the local
stability and the global selection amongst these ground states.  The
stability issue is quite delicate, since at $T = 0$ the spins can 
smoothly distort from one ground state to any other at no energy cost.  
More formally, for any ground state at $T = 0$ there is a
branch of normal modes whose frequencies $\omega_0({\bf q})$ have an infinite
number of zeros, vanishing for any ${\bf q}$ on the spiral surface.  
This leads to a divergence in a na\"ive low-temperature expansion in
small fluctuations.  To illustrate, let us start from an
arbitrary ground state ordered at wavevector ${\bf Q}$, with a
corresponding spin
configuration $\overline{{\bf S}}_i$, and expand the Hamiltonian in 
fluctuations $\delta {\bf S}_i ={\bf S}_i - \overline{{\bf S}}_i$.  To leading
order in temperature the thermally averaged fluctuation amplitude, by
equipartition, is given by
\begin{equation}
  \langle \delta{\bf S}_i^2\rangle \sim T\int \frac{d^3{\bf
  q}}{\omega^2_0({\bf q})}
  \rightarrow \infty,
  \label{corr}
\end{equation}
which diverges due to the infinite number of zeros in
$\omega_0({\bf q})$.   However, since only a finite number of these
zero-frequency modes, the ``Goldstone modes'', are
guaranteed by symmetry, thermal fluctuations can lift the remaining
``accidental'' zeros, potentially stabilizing an ordered state.

This stabilization indeed occurs.  Interestingly,
modifications to $\omega_0({\bf q})$ by thermal fluctuations are
non-perturbative in temperature.  We therefore obtain the leading corrections
for $T \ll \Theta_{CW}$ within a self-consistent treatment as described
in the Supplementary Material.  Provided $J_1 \neq 0$, we find that for
${\bf q}$ on the spiral surface the frequencies become
\begin{equation}
  \omega^2_T({\bf q}) = \omega^2_0({\bf q}) + T^{2/3}\Sigma({\bf q}),
  \label{omega}
\end{equation}
where $\Sigma({\bf q})$ is temperature-independent and generically
vanishes \emph{only} at the spiral wavevectors $\pm {\bf Q}$, which
are precisely the locations of the Goldstone modes.
Thus entropy indeed lifts the surface degeneracy, which
cures the divergence in Eq.\ (\ref{corr}) and stabilizes long-range
order.  Nevertheless, the order is in a sense ``unconventional'' in
that anomalies in thermodynamic quantities appear due to the
non-analytic temperature dependence in Eq.\ (\ref{omega}).  In
particular, the classical specific heat at low temperatures scales as
\begin{equation}
  C^{\rm classical}_v(T) = A + B T^{1/3},
  \label{specificheat}
\end{equation}
with $A$ and $B$ constants.  A crude quantum treatment, obtaining the
magnon spectrum $\epsilon({\bf q})=\hbar\omega_T({\bf q})$ by
quantizing the classical modes of Eq.\eqref{omega}, predicts the
fractional power-law $C^{\rm quantum}_v(T) \sim T^{7/3}$.  This is
intriguingly reminiscent of the approximately $T^{2.5}$ behavior
observed in \CoAl\cite{Suzuki:06} and related materials \cite{Loidl:prl04}.

\begin{figure} 
  \begin{center}
	\subfigure[]{
	  \label{fig:Numerics_1}
    \includegraphics[height=2.6in]{Tc.eps}}    
	\subfigure[]{
	  \label{fig:Numerics_2}
    \includegraphics[height=2.6in]{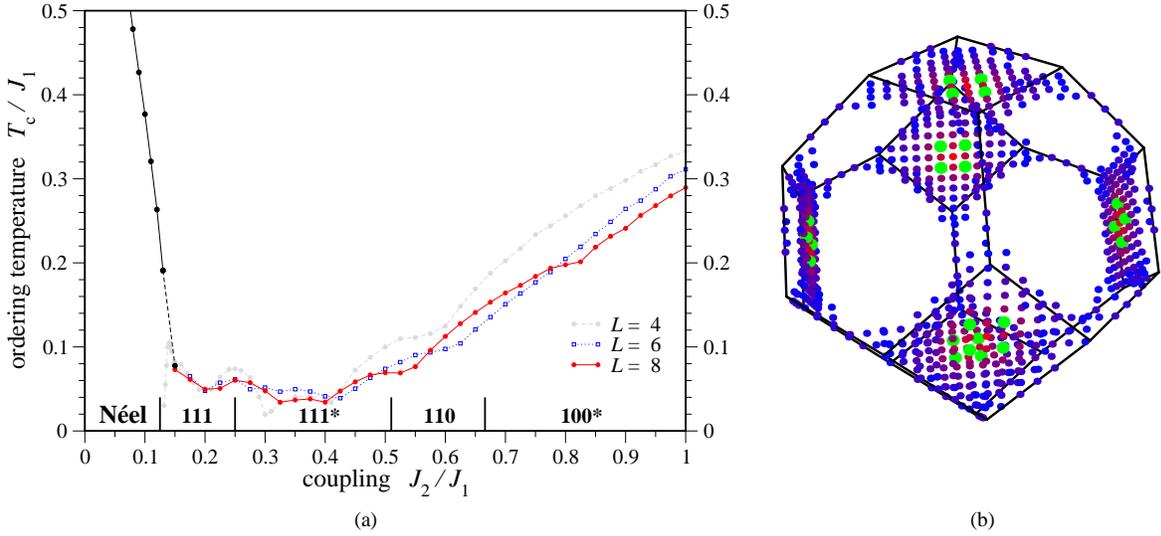}}
  \end{center} 
  \caption{
  (a) Numerical results for the ordering temperature $T_c$ versus 
  the coupling strength $J_2/J_1$ for systems with up 
  to $N=8 \times L^3=4096$ spins.  The ordering temperature rapidly diminishes 
  in the \Neel phase upon adding frustration via $J_2$, and, 
  significantly, remains finite for $J_2/J_1 > 1/8$ where
  the spiral surface occurs in agreement with our order-by-disorder
  analysis.  The entropically selected ordering at
  low-temperatures is denoted along the horizontal axis; $111^*$ and
  $100^*$ refer respectively to the green points in Figs.\
  \ref{surfaces}(b) and (c).
  The ``bumpy'' modulations in $T_c$ originate from an 
  unusual finite size effect, namely variations in the number of 
  momenta in the Brillouin zone that for the finite system
  approximate the 
  spiral surface.  
  (b) Regions of high-intensity in the magnetic
  structure factor in the paramagnetic phase just above $T_c$.  The
  data were obtained numerically for a system with
  $N = 8 \times 12^3 = 13824$ spins at coupling strength $J_2/J_1 = 0.85$
  appropriate for \MnScS.  
  As evidenced by the remarkable similarity to Fig.\
  \ref{surfaces}(c), the structure factor not only clearly reveals the
  underlying spiral surface, but also reflects the entropic corrections 
  to the free energy along the surface.
  } 
  \label{Tc} 
\end{figure}

We now address \emph{which} state thermal fluctuations select.  Although
the energy $E$ associated with each wavevector on the spiral surface is
identical, their entropy $S$ and hence \emph{free energy} $F = E-TS$
generally differ.  Typically, entropy favors states with the highest
density of nearby low-energy excitations.  To compute the free energy at
low temperatures, it suffices to retain terms in the Hamiltonian which
are quadratic in fluctuations about a state ordered at wavevector ${\bf
  Q}$.  The free energy can then be computed numerically for each ${\bf
  Q}$ on the surface.  The results for select $J_2/J_1$ are illustrated
in Fig.\ \ref{surfaces}, where the surface is colored according to the
magnitude of the free energy (blue is high, red is low, and the global
minima are green).  As indicated in Fig. \ref{fig:Numerics_1}, the free
energy minima occur at the following locations as $J_2/J_1$ varies: (i)
along the $(q,q,q)$ directions for $1/8<J_2/J_1\leq 1/4$ as in Fig.\
\ref{surfaces}(a); (ii) at the six wavevectors depicted in Fig.\
\ref{surfaces}(b) located around each ``hole'' in the surface for $1/4 <
J_2/J_1 \lesssim 1/2$; (iii) along the $(q,q,0)$ directions when $1/2
\lesssim J_2/J_1 \lesssim 2/3$; and (iv) at four points centered around
each $(q,0,0)$ direction as in Fig.\ \ref{surfaces}(c) for larger $J_2$.
Eventually the latter points converge precisely onto the $(q,0,0)$
directions, where the nearest-neighbor FCC antiferromagnet is known to
order \cite{Zhit:05}.


We next turn to the evolution with increasing temperature, for which we
rely on extensive Monte Carlo simulations and analytic
arguments.  
As one introduces frustration via $J_2$, it is natural to expect a
sharply reduced transition temperature $T_c$ relative to
$\Theta_{CW}$,  and
this is indeed borne out in our simulations.  Figure \ref{Tc}~(a)
illustrates $T_c$ versus $J_2/J_1$ computed
numerically for systems with up to $N = 4096 = 8 \times 8^3$ spins.
In the \Neel phase, a sharp decrease in $T_c$ is evident upon
increasing $J_2$.  As an interesting aside, for $J_2/J_1$ just above 1/8 
\emph{two} ordering
transitions appear below the paramagnetic phase.  This occurs due to
thermal stabilization of the \Neel phase slightly beyond the value of
$J_2/J_1 = 1/8$; the reentrant \Neel order appears below the dashed black 
line in Fig.\ \ref{Tc}(a).  More interestingly, 
$T_c$ clearly remains non-zero
for $J_2/J_1>1/8$, in agreement with the preceding order-by-disorder
analysis.  Throughout this region, the transition is strongly
first order.  

\begin{figure}
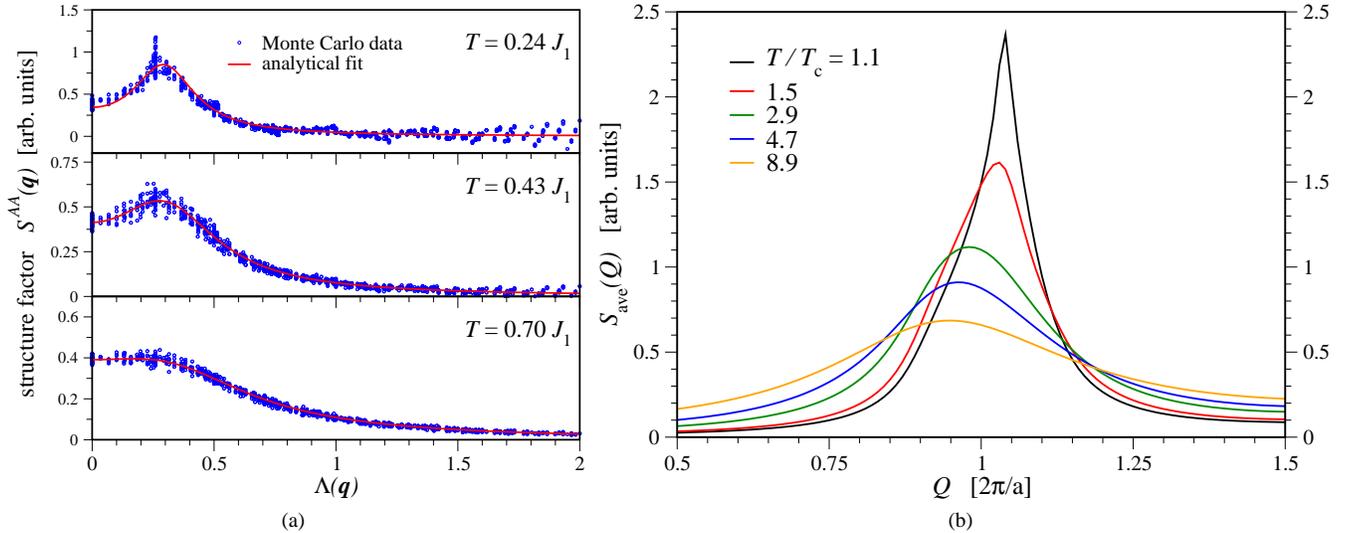
 
  \begin{center}
	\subfigure[]{
	  \label{fig:Lambda}
    \includegraphics[height=2.6in]{Lambda.eps}}    
	\subfigure[]{
	  \label{fig:Powder}
    \includegraphics[height=2.6in]{MnScS.eps}}
  \end{center} 
  \caption{(a) Structure factor data $S^{AA}({\bf q})$
  versus $\Lambda({\bf q})$ in the paramagnetic phase 
  with $J_2/J_1 = 0.85$, for which $T_c \approx 0.22 J_1$.  
  Essentially for all $T> T_c$, the numerical 
  data agree quantitatively with the spherical model predictions 
  (red curves), with one fitting parameter corresponding to an 
  overall scaling factor.  The peaks in the upper two panels
  correspond to points near the spiral surface, which remains
  discernible up to $T \approx 3 T_c$.  
  (b) Powder-averaged structure factor in the spherical
  model with $J_2 = -0.85 J_1$ and $J_3 = -J_1/20$.  The data
  correspond to temperatures ranging from just above $T_c$ (black
  curve) to several times $T_c$ (orange curve).  Corrections due to $J_3$
  initially dominate the signal, but are rapidly washed out as
  temperature increases, leaving the robust spiral spin liquid
  correlations.  The data reproduce well the diffuse
  scattering around $Q_{\rm diff} \approx 2\pi$ observed in powder
  neutron experiments.  
  } 
  \label{StructureFactorData} 
\end{figure}

Due to the strong suppression of $T_c$ when $J_2/J_1 >1/8$, one can
explore a broad range of the spin liquid regime in the paramagnetic
state above $T_c$ and below $|\Theta_{CW}|$.  Interestingly, the spiral
surface, as well as entropic free-energy corrections, can be directly
probed via the spin structure factor.  This is illustrated in Fig.\
\ref{Tc}(b), which displays the structure factor $S^{AA}({\bf q})$
corresponding to spin correlations on one of the two FCC sublattices.
(Experimentally, $S^{AA}({\bf q})$ can be obtained from the full
structure factor as described in the Supplementary Material.)  The data
correspond to $N = 13824$ spins with $J_2/J_1 = 0.85$, relevant for
\MnScS as discussed below, at a temperature just above $T_c$.  Here we
plot only momenta contributing the highest 44\% intensity (blue points
have lower intensity, red higher, and green corresponds to the maxima);
the similarity to Fig.\ \ref{surfaces}(c) is rather striking.  The
free-energy splitting manifest here persists up to $T\approx 1.3 T_c$,
while the surface itself remains discernible out to $T \approx 3 T_c$
(see Fig.\ref{fig:Lambda}).  The spiral ground states evidently dominate
the physics for $T_c \lesssim T \lesssim 3T_c$, so that this regime can
be appropriately characterized as a ``spiral spin liquid''.

To quantify the behavior in this regime analytically, we 
calculated the structure factor within the
``spherical'' approximation, in which the unit-magnitude constraint on
each spin is relaxed to $\sum_i |{\bf S}_i|^2=N$.  The classical spin liquids
in kagome \cite{PhysRevB.59.443} and pyrochlore \cite{isakov:167204}
antiferromagnets are known to be well-described by this scheme.  
Here, we find that the structure factor is similarly peaked on the spiral
surface, with a width $\xi^{-1}\sim k_{\scriptscriptstyle B}T$ that
agrees quantitatively with the fitted value from numerics.  
Moreover, the complete three-dimensional structure factor data collapse
onto a (known) one-dimensional curve when plotted versus the
variable $\Lambda({\bf q}) = 2\sqrt{\cos^2 \frac{q_x}{4}\cos^2
  \frac{q_y}{4} \cos^2 \frac{q_z}{4}+\sin^2 \frac{q_x}{4}\sin^2
  \frac{q_y}{4} \sin^2 \frac{q_z}{4}}$.
As illustrated in Fig.\ \ref{StructureFactorData}(a) 
for $J_2/J_1 = 0.85$, the numerical data conform well to this 
prediction, essentially throughout the paramagnetic phase
except very near $T_c$ where thermal fluctuations dramatically split
the free energy along the surface.  Note that the red
analytical curves contain only a single fitting parameter,
corresponding to an unimportant overall scaling.


Finally, we discuss implications
for experiments, focusing on the well-characterized material \MnScS.  
Below $T_{N1} = 2.3$ K, experiments observe
long-range spiral order with wavevector ${\bf Q}_{exp} \approx
2\pi(3/4,3/4,0)$, coexisting with pronounced correlations with
wavevector magnitude $Q_{\rm diff} \approx 2\pi$  
that persist to well above $T_{N1}$ \cite{Loidl:prb06,Mucksch:06}.  A second
transition occurs at $T_{N2} = 1.9$ K, below which the latter
correlations are greatly suppressed.  
Assuming ${\bf Q}_{exp}$ lies near the
spiral surface, we estimate that $J_2/J_1 \approx 0.85$ for this
material.  By comparing the structure of the spiral ground state ordered
at ${\bf Q}_{exp}$ with the experimentally determined spin structure
\cite{Loidl:prb06}, we further deduce that $J_1$ must be ferromagnetic
(\emph{i.e.}, $J_1<0$) in \MnScS.  One can then extract the
magnitudes of the exchange constants from the measured Curie-Weiss
temperature $\Theta_{CW} \approx -22.1$ K; we obtain $J_1 \approx -1.2$ K
and $J_2 \approx 1.0$ K.  According to the numerical results of Fig.\
\ref{Tc}, the predicted ordering temperature for these parameters is
$T_c \approx 2.4$K.

Obtaining a detailed comparison to the \emph{low-temperature} experimental
order requires perturbing our parent Hamiltonian.  
Quite generally, these corrections inevitably
overwhelm the entropic free energy splittings discussed above at
sufficiently low temperature, since the latter vanish as $T\rightarrow
0$.  Happily, the simplest correction---a small antiferromagnetic
third-neighbor exchange $J_3$---favors the observed $(q,q,0)$ spiral
direction.  The close proximity of the calculated $T_c$ for $J_3=0$ to
the experimental $T_{N1}$ suggests that $J_3$ should indeed be
small.  For sufficiently small $J_3$, the entropic splittings will
outweigh this energetic correction at higher temperatures, giving way
to an intermediate phase with long-range spiral order along the
entropically favored (approximately) $(q,0,0)$ directions.  As $J_3$
increases (but remains small), this order-by-disorder phase will be weakened
and eventually removed, leaving only the more robust spiral spin liquid
correlations above $T_c$.  The $Q_{\rm diff}$ scattering deduced from powder
neutron experiments is consistent with weak order-by-disorder as well
our predictions for the spiral spin liquid, and further 
single-crystal experiments are needed to distinguish between these scenarios.  
For comparison with the latter, in Fig.\ \ref{StructureFactorData}(b) 
we display the numerically powder
averaged spherical model structure factor $S_{\rm ave}(Q)$ for 
$J_3 = |J_1|/20$ at several temperatures above $T_c$.  
This reproduces well the experimental diffuse
correlations near $Q_{\rm diff}$ as temperature smears the $J_3$ splitting.  
In short, the ``competing order'' observed at intermediate
temperatures is precisely in line with theoretical expectations in our
framework, and thus in our view provides convincing experimental
evidence for our model's relevance to the
physics of \MnScS.


Looking forward, many other materials are anticipated to be
well-described by our model, from the marginally frustrated
CoRh$_2$O$_4$ and MnAl$_2$O$_4$ with $f \approx 1.2$ and $f \approx
3.6$, respectively, to the highly frustrated CoAl$_2$O$_4$.  Existing
measurements place a lower bound on the frustration parameter for
CoAl$_2$O$_4$ of around 10-20; a broad
peak in the specific heat evidently preempts a sharp
ordering transition in current samples \cite{Loidl:prb05,Suzuki:06}.  
As described in the Supplementary Material, the available 
low-temperature powder neutron data together with the large
frustration parameter are consistent with this material residing in
the region $J_2/J_1 \approx 1/8$, where the spiral surface begins to
develop.  Experiments with increased sample purity would likely allow
for a more direct comparison.  Regarding future experiments more
generally, most exciting would be single crystal neutron data, which
would allow a much more direct and detailed comparison of theory and
experiment.  One could carry out an analysis similar to the one 
performed for the structure factor in our Monte Carlo simulations,
as detailed in the Supplementary Material.  In this way, one might
directly observe the spiral surface in the spiral spin liquid regime
and perhaps find the first unambiguous experimental signatures of 
order-by-disorder, both of which would be truly remarkable.


\begin{acknowledgments}  
  We would like to acknowledge Ryuichi Shindou, Zhenghan Wang, and
  Matthew Fisher for illuminating discussions, as well as Tomoyuki
  Suzuki, Michael Muecksch, and Alexander Krimmel for sharing
  their unpublished results.
  This work was supported by the Packard Foundation (D.\ B.\ and L.\
  B.) and the National Science Foundation 
  through grants DMR-0529399 (J.\ A.), and DMR04-57440 (D.\ B.\ and L.\
  B.).     
\end{acknowledgments}  


\begin{center}
    {\Large {\bf Supplementary Material}}
\end{center}

\section{Low temperature}

\subsection{Ground states}

To find the ground states of our parent theory, it is useful to
diagonalize the Hamiltonian by transforming to momentum space. 
Since the diamond lattice is an FCC Bravais lattice with a two-site
basis, this reveals two bands with energies
\be
  \epsilon_{\pm}({\bf q}) = 4 J_2 [\Lambda^2({\bf q})-1] \pm 2 
  J_1 \Lambda({\bf q}).
\ee
Here and below we assume antiferromagnetic $J_{1,2} > 0$; the 
function $\Lambda({\bf q})$ is defined by
\begin{eqnarray}
  \Lambda({\bf q}) &=& 2 [\cos^2(q_x/4)\cos^2(q_y/4)\cos^2(q_z/4)
  \nonumber \\
  &+& \sin^2(q_x/4)\sin^2(q_y/4)\sin^2(q_z/4)]^{1/2}
\end{eqnarray}
as provided in the main text.  The minimum eigenvalue is realized in
the lower band $\epsilon_-({\bf q})$, and occurs at a single point
(${\bf q = 0}$) for $J_2/J_1 < 1/8$ but on a two-dimensional
\emph{surface} in momentum space for larger $J_2/J_1$.  

Our strategy is to explicitly construct states that contain Fourier 
weight only at the minimum eigenvalue, \emph{and} simultaneously 
satisfy the unit-vector constraint for each spin.  Such states are
guaranteed to be ground states, as mixing any other Fourier components
necessarily increases the energy.  For $J_2/J_1 < 1/8$ this
prescription gives the expected \Neel phase as the unique ground state
(apart from global spin rotations).  For larger $J_2/J_1$ one can
construct highly degenerate spiral ground states, each characterized
by a single wavevector lying on the ``spiral surface''
corresponding to the minimum of $\epsilon_-({\bf q})$.  Denoting the
two FCC sublattices by A and B and the lattice site positions by ${\bf r}_j$, 
the spiral ground states explicitly take the form
\begin{eqnarray}
  {\bf S}^{A/B}_j &=& \mp [{\bf \hat x} \cos \varphi^{A/B}_j 
  + {\bf \hat y}\sin \varphi^{A/B}_j]
  \\
  \varphi^{A/B}_j &=& {\bf q}\cdot{\bf r}_j\pm \theta({\bf q})/2,
\end{eqnarray}
with any wavevector ${\bf q}$ on the spiral surface.  We have assumed
a spiral in the $x$-$y$ plane, though any two orthonormal unit vectors
above will clearly do.  The angle $\theta({\bf q})$ determines the
relative phase shift between the A and B sublattices, and is given by
the argument of 
\begin{equation}
  \cos{\left(\frac{q_1}{4}\right)} \cos{\left(\frac{q_2}{4}\right)} 
  \cos{\left(\frac{q_3}{4}\right)} - 
  i\sin{\left(\frac{q_1}{4}\right)} \sin{\left(\frac{q_2}{4}\right)} 
  \sin{\left(\frac{q_3}{4}\right)} \; .
\end{equation}
(Note that for ferromagnetic $J_1 < 0$, the corresponding
ground states are obtained by reversing the spins on one FCC
sublattice.)  

While this does not exhaust all possible ground states, others occur
only at special values of $J_2/J_1$ or contribute only a finite
discrete set and are thus anticipated to be less important than these
generic spirals.  For instance, a discrete set of ground states
constructed from wavevectors on the surface differing by half a
reciprocal lattice vector can be realized over a range of $J_2/J_1$.

\subsection{Local stability}

Henceforth we focus on the regime $J_2/J_1 >1/8$.  Given the massive
spiral ground state degeneracy here, the question of stability of
long-range order becomes quite delicate.  The goal of this subsection is
to demonstrate that \emph{entropy} stabilizes long-range order at
finite temperature by lifting the degeneracy in the free energy along
the spiral surface, \emph{i.e.}, the system undergoes a thermal
order-by-disorder transition.

To this end, we start from an arbitrary ground state ordered at
momentum ${\bf Q}$ with a spin
configuration $\overline {\bf S}_j$ and expand in fluctuations
by writing
\be
  {\bf S}_j = {\vec \pi}_j + \overline{{\bf S}}_j
  \sqrt{1 - {\vec \pi}^2_j} \; .
\ee
The fluctuation field $\vec \pi_j$ is constrained such that $\overline
{\bf S}_j\cdot \vec \pi_j = 0$ so that the unit-vector
constraint remains satisfied.  After computing the Jacobian 
for the variable transformation, the partition function becomes
\be
\begin{split}
{\mathcal Z} = & \int {\mathcal D}{\bf S} e^{-\beta {H}}
\prod_{\bf r} \delta[{\bf S}^2_j - 1] 
\\ 
= & \int {\mathcal D}{\vec \pi} e^{-\beta {H}} \prod_{\bf
  r}  
  \left[ 1 - {\vec \pi}^2_j \right]^{-1/2}
\; .
\end{split}
\ee
An expansion in small fluctuations can be controlled at low temperatures.  
Assuming the spins $\overline {\bf S}_j$ lie in the $x$-$y$
plane, we parametrize the fluctuations as follows,
\be
  {\vec \pi}_j = {\bf \hat z} \phi_j + 
  [{\bf \hat z} \times \overline{{\bf S}}_j ]
  \chi_j \; ,
  \label{parametrization}
\ee
thereby automatically satisfying the 
constraint $\overline {\bf S}_j\cdot \vec \pi_j = 0$.
The partition function can now be expressed in terms of an action,
\be
  {\mathcal Z} = \int {\mathcal D}\phi {\mathcal D}\chi e^{-S}
\; .
\ee
Retaining the leading corrections to the Gaussian theory, 
the action can be written as $S = S_2+S_3+S_4$, where 
\be\label{action}
\begin{split} &
{\mathcal S}_2 = 
\frac{\beta}{2} \sum_{i j} 
\left[ 
{\tilde J}_{i j}
\phi_i \phi_j + W_{i j} \chi_i \chi_j
\right] 
-\frac{1}{2}\sum_j[\phi_j^2 + \chi_j^2]
\\ &
{\mathcal S}_3 = 
\frac{\beta}{2} \sum_{i j} K_{i j} 
\phi_i 
\left( \phi_j^2 + \chi_j^2 \right)
\\ &
{\mathcal S}_4 = 
\frac{\beta}{8} \sum_{i j} W_{i j} 
\left( \phi_i^2 + \chi_i^2 \right)
\left( \phi_j^2 + \chi_j^2 \right)
\; .
\end{split}
\ee
Here ${\tilde J}_{i j}$ is simply the exchange matrix $J_{i j}$ 
shifted by a constant, such that all the eigenvalues are non-negative and
the ground state space corresponds to the kernel of this matrix.
We have also defined the matrices 
$W_{i j} = {\tilde J}_{i j} \left( \overline{{\bf S}}_j \cdot \overline{{\bf S}}_i \right)$
and
$K_{i j} = {\tilde J}_{i j}
\left[ 
{\hat z} \cdot \left( \overline{{\bf S}}_j \times \overline{{\bf S}}_i \right)
\right]$.  The Jacobian factor has been absorbed into the action,
giving rise to the last summation in $S_2$.  

According to Eq.\ (\ref{parametrization}), fluctuations out of 
the spiral plane are described by $\phi_j$, while
$\chi_j$ describes in-plane fluctuations.  Long-range order will occur
if these fluctuations can always be made small by going to
sufficiently low temperature.  The latter generically have
only a single gapless mode, corresponding to the symmetry-required
Goldstone mode at zero momentum.  Consequently, fluctuations in
$\chi_j$ are clearly well-behaved at low temperature.  Subtleties with
long-range order arise from the $\phi_j$ fluctuations, which connect
the degenerate ground states.  At the Gaussian level and to leading
order in temperature, the $\phi_j$ propagator is
\be
  G_{ij}^0 = \langle \phi_j \phi_i \rangle_0 = {\tilde J}_{i j}^{-1}.
\ee
In momentum space, the associated normal mode frequencies
$\omega_0({\bf q})$ and $\omega_1({\bf q})$ are defined by
\begin{equation}
  \omega_{0,1}^2({\bf q}) \equiv \epsilon_\mp({\bf q})-\epsilon^{\rm
  min}_-, 
\end{equation}
where $\epsilon^{\rm min}_-$ corresponds to the minimum
value of $\epsilon_-({\bf q})$.  
It follows that the fluctuation amplitude for 
$\phi_j$ naively \emph{diverges},
\be
\langle \phi_j^2 \rangle_0 \sim T \int_{\bf q} \frac{1}{\omega_0^2({\bf q}) }
\rightarrow \infty 
\; ,
\ee
since $\omega_0({\bf q})$ vanishes for any ${\bf q}$ on the spiral
surface due to the continuous ground state degeneracy. 

Higher-order corrections in temperature, however, lift the surface
degeneracy, thus curing the above divergence and stabilizing
long-range order.  Perturbation theory
in temperature suffers similar divergences as found above, and hence 
we employ a self-consistent treatment to obtain corrections to the
$\phi_j$ fluctuations.  The $\phi_j$ propagator obtained from the full
action $S$ defined above is 
\be
  G_{ij} = \langle \phi_j \phi_i \rangle = \left[ {\tilde J}_{i j} + {\tilde
    \Sigma}_{i j} \right]^{-1},
\ee
where $\tilde \Sigma_{ij}$ is the \emph{self-energy}.  In particular,
we are interested in the self-energy correction to $\omega_0({\bf
  p})$, which we will denote $\tilde \Sigma({\bf p})$, for momenta
${\bf p}$ along the spiral surface.  

To proceed, we first assume that thermal
fluctuations indeed break the surface degeneracy, and then find the leading
corrections self-consistently.  More specifically, we assume that 
${\tilde \Sigma}({\bf p}) \sim T^\alpha \Sigma({\bf p})$, where
$\alpha <1$, $\Sigma({\bf p})$ is temperature-independent, 
and $\Sigma(\pm{\bf Q}) = 0$; the last condition simply
asserts that the symmetry-required Goldstone modes at the ordering
wavevectors are preserved.  With these assumptions, we obtain a 
self-consistent equation of the form
\be\label{sigma}
{\tilde \Sigma}({\bf k}) =
 T \int_{\bf q} 
\Gamma({\bf q},{\bf k})
G({\bf q})
\; ,
\ee
with $G({\bf q}) = \left[ \omega_0^2({\bf q}) 
+ {\tilde \Sigma}({\bf q}) \right]^{-1}$ and $\Gamma({\bf q},{\bf k})$
temperature-independent.  
At low-temperatures, the integral is dominated by momenta near the
spiral surface due to the propagator $G({\bf q})$.  By contrast, the
function $\Gamma({\bf q}, {\bf k})$ is well behaved and does not lead to any 
additional singular behavior.  Hence it is sufficient to replace
$\Gamma({\bf q},{\bf k})\rightarrow \Gamma({\bf q}_s,{\bf k})$ 
under the integral, where ${\bf q}_s$ lies precisely on the surface in
the direction of ${\bf q}$.  One can show that $\Gamma({\bf q}_s,\pm
{\bf Q}) = 0$, so that the Goldstone modes are indeed preserved within
our self-consistent treatment.  Furthermore, one can approximate
$G({\bf q}) \approx \left[ \kappa \left(q - q_s\right)^2 
+ T^\alpha \Sigma({\bf q}) \right]^{-1}$ in the integrand.
The temperature dependence can then be scaled out of the integral,
implying a power $\alpha = 2/3$ consistent with our assumptions.

The divergent fluctuations are thus cured by the onset of a thermally
induced splitting $\Delta \sim T^{2/3}$ along the spiral surface.
Consequently, ordering at finite temperature will occur,
despite the massive ground state degeneracy.

\subsection{Global selection}

\begin{figure} 
  \begin{center} 
    \includegraphics[height = 2.0in]{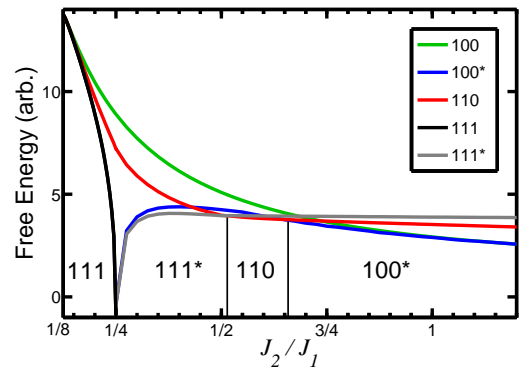}
  \end{center} 
  \caption{Free energy versus $J_2/J_1$ along high-symmetry directions
    in the Brillouin zone.  
} 
  \label{FreeEnergy} 
\end{figure}

In the previous subsection we found that thermal fluctuations
stabilize long-range order at finite temperature.  Here we address the
more specific (and simpler) question of which state among the 
degenerate set is favored.  At finite temperature, entropy selects 
the states minimizing the free energy $F = E-TS$ ($E$ is energy, $S$
entropy), which usually are
those with the highest density of nearby low-energy states.  
Let us start from an arbitrary spiral with ordering wavevector 
${\bf Q}$, and expand in fluctuations as outlined in the previous
subsection.  At low temperatures, for our purpose here it suffices to
retain only the first two terms in the Gaussian action $S_2$.
Integrating over the fluctuation fields, we then obtain the leading
$T$- and ${\bf Q}$-dependent contribution to the free energy,
\be\label{freeE}
\begin{split}
F({\bf Q}) & = - T \ln({\mathcal Z}) 
\\ & \sim
T\, \left(
 Tr\left[ 
\ln({\tilde J}/2\pi T)
\right]
+ Tr\left[ 
\ln(W({\bf Q})/2\pi T)
\right]
\right)
\; ,
\end{split}
\ee
where ${\hat J}, {\hat W}$ denote the matrices defined in the 
previous subsection and we have explicitly labeled the ${\bf
  Q}$-dependence in ${\hat W}$.  The first term is ${\bf Q}$-independent
and thus does not distinguish the states on the spiral surface.  This
is accomplished, however, by the second term, which can be easily
computed numerically as a function of ${\bf Q}$ to obtain the global free
energy minima.  The resultant free-energy splittings are illustrated
through the coloring of the surfaces in Fig. 2 of main text.  Also,
we display in Fig.\ \ref{FreeEnergy} the free energy along
high-symmetry directions as a function of $J_2/J_1$.  Here, $111^*$
refers to six momenta located along the ``holes'' which develop in the
surface for $J_2/J_1>1/4$, and $100^*$ corresponds to four momenta
located around the $100$ directions (see main text).  

\subsection{Specific heat}

The anomalous low temperature dependence of the free energy will manifest itself in thermodynamic quantities.
In this section we show explicitly how the specific heat varies with temperature in this regime.

The heat capacity $ C_v = - T \left( \frac{\partial^2 F}{\partial T^2} \right)_{V,N} $ can be found from our
low temperature expression for the free energy. Including only the anomalous part of the self energy,
while neglecting all analytic corrections higher order in $T$,
we modify \eqref{freeE} to
\be
\begin{split}
F \sim &
T\, \left(
 Tr\left[ 
\ln[({\tilde J} + {\tilde \Sigma})/2\pi T]
\right]
+ Tr\left[ 
\ln(W({\bf Q})/2\pi T)
\right]
\right)
\\
\sim &
- A_1 T \ln(T)
+ A_2 T
+ T \int_{\bf q}
\ln[(\omega_0^2({\bf q}) + T^{2/3} \Sigma({\bf q}) )]
\; .
\end{split}
\ee

To find the behavior of the integral at low temperatures, it is useful to consider
\be
\frac{\partial (F/T)}{\partial T} \sim
- \frac{A_1}{T} + \frac{2}{3} T^{-1/3} \int_{\bf q}
\frac{\Sigma({\bf q})}{(\omega_0^2({\bf q}) + T^{2/3} \Sigma({\bf q}) )]}
\ee
scaling temperature out of the momentum integral on the right hand side. In the same manner
we proceeded for the integral in \eqref{sigma}, we find 
$\int_{\bf q}
\frac{\Sigma({\bf q})}{(\omega_0^2({\bf q}) + T^{2/3} \Sigma({\bf q}) )]} \sim T^{-1/3}$
so that
$
\frac{\partial (F/T)}{\partial T} \sim
- \frac{A_1}{T} + \frac{2}{3} T^{-2/3} B
$. The low temperature form of the free energy is
\be
F \sim - A_1 T \ln(T) + A_2 T + A_3 T^{4/3}
\ee
where $A_{1,2,3}$ are constants.
$ $From this form it follows that the heat capacity is
\be
C^{\rm classical}_v(T) = A + B T^{1/3}
\; .
\ee

\section{High temperature}

This section is concerned with analytically 
describing the spin correlations at temperatures above $T_c$.  
Remarkably, these allow one to probe directly the underlying 
ground state surface in the ``spiral spin liquid'' regime occurring 
over a broad temperature range.  In the disordered phase above $T_c$, the 
spins fluctuate strongly, and it is reasonable that the unit length 
constraint on the individual spins can be relaxed.  Hence we employ
the ``spherical'' approximation, replacing the local unit-vector spin
constraint with the global constraint $\sum_j {\bf S}_j^2 = N$, 
$N$ being the total number of sites.  The spin correlations determined
via Monte Carlo numerics are described \emph{quantitatively} within 
this scheme, except very near $T_c$ where entropic effects are dramatic.  

The partition function for this model is 
\be
  {\mathcal Z} = \int {\mathcal D}{\bf S}d\lambda e^{-\beta {H}
  - i \lambda (\sum _j {\bf S}_j^2 - N)}
\; ,
\ee
where $\lambda$ is a Lagrange multiplier enforcing the global constraint.  
To proceed we employ a saddle-point approximation, replacing $i\lambda 
\rightarrow \beta \Delta(T)/2$, where $\Delta(T)$ is the saddle-point
value to be determined.  The spin correlation function is then
\be
\langle {\bf S}_i {\bf S}_j \rangle = 3 T \left[
J_{i j} + \delta_{i j} \Delta(T)
\right]^{-1}
\; .
  \label{correlation}
\ee
Upon integrating over the spins, one obtains the saddle point equation
for $\Delta(T)$:
\be
  \frac{1}{T} = \frac{3}{2} \int_{\bf q} \sum_{j = 0,1} 
  \frac{1}{\omega_j^2({\bf q}) + \Delta(T)}
\; .
  \label{sp}
\ee

Equations (\ref{correlation}) and (\ref{sp}) together determine the
spherical model spin correlations.  In particular, 
the momentum-space correlation function for spins on the same 
FCC sublattice is given by
\begin{equation}
  S^{AA}({\bf q}) \sim T \bigg{[}\frac{1}{\omega_0^2({\bf q}) + 
  \Delta(T)} + \frac{1}{\omega_1^2({\bf q}) + 
  \Delta(T)} \bigg{]},
  \label{SAA}
\end{equation}
while the correlation between spins on opposite FCC sublattices is
\begin{equation}
  S^{AB}({\bf q}) \sim T e^{-i\theta({\bf q})}
  \bigg{[}\frac{-1}{\omega_0^2({\bf q}) + 
  \Delta(T)} + \frac{1}{\omega_1^2({\bf q}) + 
  \Delta(T)} \bigg{]}.
\end{equation}
The full structure factor, as measured in experiment, is 
\begin{equation}
  S({\bf q}) = S^{AA}({\bf q}) + \text{Re}[S^{AB}({\bf q})].
\end{equation}

Notice that the spin correlation $S^{AA}({\bf q})$ depends 
on momentum only through the function $\Lambda({\bf q})$.  (${\bf
  S}^{AB}({\bf q})$ has additional momentum dependence through
$\theta({\bf q})$.)  Hence it is highly desirable to isolate this
contribution, as $S^{AA}({\bf q})$ collapses onto a known one-dimensional curve
when plotted versus $\Lambda({\bf q})$.  
We have extracted $S^{AA}({\bf q})$ in our Monte Carlo simulations 
for $J_2/J_1 = 0.2,0.25,0.4,0.6,0.85$, and indeed find that in all 
cases for $T > T_c$ the correlation function data collapse well when
plotted versus $\Lambda({\bf q})$.  Furthermore, in all these cases
one finds \emph{quantitative} agreement with 
the analytic result Eq.\ (\ref{SAA}), with only a single fitting
parameter corresponding to an overall scaling.  The excellent
agreement obtained here is illustrated in the main text for $J_2/J_1 =
0.85$.  The peaks in these figures correspond to values of
$\Lambda({\bf q})$ defining the spiral surface, thus implying that
spin configurations near the surface dominate the physics.
This is the spiral spin liquid regime.

Naively, isolating $S^{AA}({\bf q})$ experimentally appears more
difficult.  Fortunately, this component
can be extracted from the full structure factor by using the fact that
$\theta({\bf q}) = \theta({\bf q}+ {\bf K})+\pi$, where ${\bf K} =
4\pi(1,0,0)$ is a reciprocal lattice vector.  This leads to the useful
identity
\begin{equation}
  S^{AA}({\bf q}) = \frac{1}{2}[S({\bf q}) + S({\bf q}+{\bf K})].
\end{equation}
It would be extremely interesting to perform a similar analysis on 
experimental neutron scattering data, which would require single 
crystals.  The spiral surface could then be extracted quite simply 
as follows.  Display momenta in the first Brillouin zone corresponding to the
highest intensity points within some threshold---the surface is 
mapped out when an appropriate threshold is
chosen.  Such an analysis was carried out for the Monte Carlo structure 
factor, the result of which are shown in the main text.

\begin{figure}
	\centering
		\includegraphics[height=2.0in]{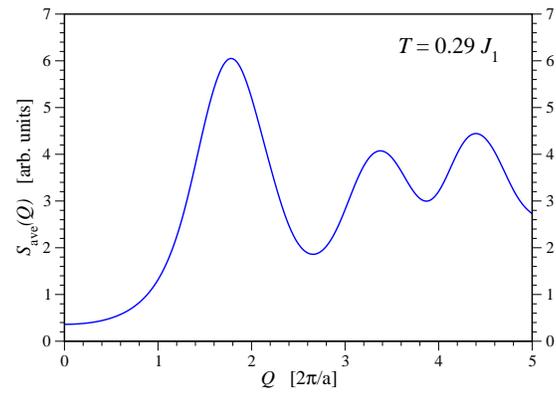}
	\caption{Powder-averaged structure factor in the spherical
  model with $J_2/J_1 \approx 1/8$.  The data reproduce well the diffuse
  scattering observed in powder
  neutron experiments.}
	\label{fig:CoAlO}
\end{figure}

Obtaining single crystal samples is often challenging, so it is highly
desirable (and of current experimental relevance) to have a way of
detecting the spiral surface in neutron data for powder samples.  The
full structure factor can be numerically ``powder-averaged'' by
performing an angular integration for a given wavevector magnitude
$Q$:
\begin{equation}
  S_{\rm ave}(Q) = \int \sin\theta d\theta d\varphi S({\bf Q}),
\end{equation}
where $\theta$ and $\varphi$ are polar and azimuthal angles specifying 
the direction of ${\bf Q}$.  The spiral surface is then indirectly revealed as a
peak in $S_{\rm ave}(Q)$ over the range of $Q$ for which the surface
occurs.  Existing neutron data for MnSc$_2$S$_4$ powder samples 
indeed reveal a broad peak in the structure factor in agreement with
our predictions for the spiral spin liquid regime.  Furthermore,
excellent agreement with powder neutron data for CoAl$_2$O$_4$ can be
obtained by assuming that $J_2/J_1 \approx 1/8$ for this material.
Fig.~\ref{fig:CoAlO} displays the predicted powder-averaged structure factor,
which exhibits peaks and valleys that correspond well to those
observed experimentally \cite{Loidl:physicaB06}.  
The low-$T_c$ in this vicinity of $J_2/J_1$ is further 
consistent with the large frustration parameter observed for 
CoAl$_2$O$_4$. 

\section{Monte Carlo methods}
In our numerical simulations of Hamiltonian we used classical
Monte Carlo techniques employing a parallel tempering scheme 
 \cite{ParallelTempering} where multiple replicas of the system are 
 simulated simultaneously over a range of temperature. 
 Thermal equilibration can be dramatically increased by swapping replicas
 between neighboring temperature points. An optimal set of temperature 
 points in the vicinity of the phase transition has been identified for each ratio
 of competing interactions $J_2/J_1$ applying a recently introduced feedback
 technique \cite{OptimizedEnsembles,OptimizedTempering}.
The implementation of these algorithms was based on the ALPS libraries \cite{ALPS}.

\bibliographystyle{naturemag}

\vfill\eject

\end{document}